\documentclass[sigconf,nonacm]{acmart}

\AtBeginDocument{%
  }

\setcopyright{none}
\copyrightyear{2018}
\acmYear{2018}
\acmDOI{XXXXXXX.XXXXXXX}

\acmConference[Conference acronym 'XX]{Make sure to enter the correct
  conference title from your rights confirmation email}{June 03--05,
  2018}{Woodstock, NY}

\acmISBN{978-1-4503-XXXX-X/2018/06}

\usepackage{algorithm}
\usepackage{algorithmic}
\usepackage{amsmath}
\usepackage{booktabs}
\usepackage{nicefrac}
\usepackage{microtype}
\usepackage{xcolor}
\usepackage{subcaption}
\usepackage{multirow}
\usepackage{makecell}
\usepackage{pifont}
\usepackage{array}
\usepackage{url}
\usepackage{tabularx}
\usepackage{enumitem}
\usepackage{colortbl}
\newcolumntype{P}[1]{>{\raggedright\arraybackslash}p{#1}}
\newcolumntype{L}[1]{>{\raggedright\arraybackslash}p{#1}}
\newcolumntype{C}[1]{>{\centering\arraybackslash}p{#1}}
\newcolumntype{Y}{>{\raggedright\arraybackslash}X}
\newcolumntype{Z}{>{\centering\arraybackslash}X}
\definecolor{lightgrey}{rgb}{0.93,0.93,0.93}
\definecolor{yellowtext}{RGB}{68,132,243}
\definecolor{yellowred}{RGB}{50,167,82}
\definecolor{yellowblue}{RGB}{251,191,5}
\definecolor{darkblue}{rgb}{0, 0, 0.5}
\definecolor{darkgreen}{rgb}{0.0, 0.42, 0.24}
\definecolor{maroon}{HTML}{A00000}
\definecolor{gray}{rgb}{0.5, 0.5, 0.5}
\definecolor{chocolate}{HTML}{D2691E}
\definecolor{indigo}{HTML}{4B0082}
\definecolor{violet}{HTML}{4B2E83}
\definecolor{lightgreen}{HTML}{E0FBE0}
\definecolor{lightred}{HTML}{FBE0E0}
\definecolor{paperorange}{HTML}{C04000}
\definecolor{lightblue}{rgb}{0.0, 0.0, 0.5}
\definecolor{cadmiumgreen}{rgb}{0.0, 0.42, 0.24}
\definecolor{forestgreen}{rgb}{0.13, 0.55, 0.13}
\definecolor{lightbluebg}{RGB}{235, 245, 255}
\definecolor{blueframe}{RGB}{70, 130, 180}
\definecolor{DeepPurple}{RGB}{103, 58, 183}
\definecolor{LightPurple}{RGB}{237, 231, 246}
\definecolor{lightgray}{rgb}{0.9, 0.9, 0.9}
 \definecolor{g100}{HTML}{2E7D32}   
\definecolor{g90}{HTML}{388E3C}    
\definecolor{g80}{HTML}{43A047}    
\definecolor{g70}{HTML}{66BB6A}    
\definecolor{g60}{HTML}{81C784}    
\definecolor{g50}{HTML}{A5D6A7}    
\definecolor{g40}{HTML}{C8E6C9}    
\definecolor{g30}{HTML}{E8F5E9}    
\definecolor{g20}{HTML}{F1F8E9}    
\definecolor{g0}{HTML}{FFFFFF}     

\definecolor{pur1}{RGB}{243, 246, 251}   
\definecolor{pur2}{RGB}{224, 231, 243}   
\definecolor{pur3}{RGB}{196, 208, 229}   
\definecolor{pur4}{RGB}{164, 183, 216}   
\definecolor{pur5}{RGB}{129, 153, 196}   
 \usepackage{xcolor}    
\usepackage{listings}  
\usepackage{float}     
\usepackage{booktabs}  
\usepackage{subcaption}
\usepackage{graphicx}
\usepackage{xcolor}
\definecolor{diffgreen}{rgb}{0.0, 0.5, 0.0} 
\definecolor{diffred}{rgb}{0.6, 0.0, 0.0}   
\definecolor{codeblue}{rgb}{0.0, 0.0, 0.8}  
\usepackage[table]{xcolor}  
\usepackage{arydshln}       
\definecolor{codegreen}{rgb}{0,0.6,0}
\definecolor{codegray}{rgb}{0.5,0.5,0.5}
\definecolor{codepurple}{rgb}{0.58,0,0.82}
\definecolor{backcolour}{rgb}{0.95,0.95,0.92}

\lstdefinelanguage{JavaScript}{
  keywords={typeof, new, true, false, catch, function, return, null, catch, switch, var, if, in, while, do, else, case, break},
  keywordstyle=\color{blue}\bfseries,
  ndkeywords={class, export, boolean, throw, implements, import, this},
  ndkeywordstyle=\color{codegreen}\bfseries,
  identifierstyle=\color{black},
  sensitive=false,
  comment=[l]{//},
  morecomment=[s]{/*}{*/},
  commentstyle=\color{codegreen}\ttfamily,
  stringstyle=\color{red}\ttfamily,
  morestring=[b]',
  morestring=[b]"
}
\lstdefinestyle{mystyle}{
    backgroundcolor=\color{backcolour},
    commentstyle=\color{codegreen},
    keywordstyle=\color{magenta},
    numberstyle=\tiny\color{codegray},
    stringstyle=\color{codepurple},
    basicstyle=\ttfamily\footnotesize, 
    breakatwhitespace=false,
    breaklines=true,
    captionpos=b,
    keepspaces=true,
    numbers=left,
    numbersep=5pt,
    showspaces=false,
    showstringspaces=false,
    showtabs=false,
    tabsize=2,
    frame=single                     
}
\usepackage[framemethod=tikz]{mdframed}
\newenvironment{custommdframed}
  {\begin{mdframed}[style=customstyle]}
  {\end{mdframed}}
 \mdfdefinestyle{customstyle}{
  linecolor=gray!100,
  linewidth=3pt,
  innerleftmargin=3pt,
  topline=false,
  rightline=false,
  bottomline=false,
  leftline=true,
  innerrightmargin=3pt,
  innertopmargin=3pt,
  innerbottommargin=3pt,
  backgroundcolor=gray!15
}

\newcommand{\pcc}[2]{\cellcolor{#1}{#2}}

\newcommand{\tabledoublesetup}{%
  \small
  \setlength{\tabcolsep}{4.5pt}%
  \renewcommand{\arraystretch}{1.15}%
}

\definecolor{crashred}{RGB}{204,0,0}
\definecolor{safegreen}{RGB}{0,153,0}
\newcommand{\TextCircle}[1][0.7]{%
    \textcolor{yellowtext}{\textbf{[T]}}%
}
\newcommand{\ImageCircle}[1][0.76]{%
    \textcolor{yellowred}{\textbf{[I]}}%
}

\newcommand{\cmark}{\ding{51}}

\newcommand{\tool}{\textsc{ContraFix}}

\begin{document}

\title{\tool{}: Skill-Enhanced Contrastive Runtime Analysis for Vulnerability Repair}

\author{Simiao Liu}
\affiliation{%
  \institution{Beihang University}
  \city{Beijing}
  \country{China}
}
\email{buaalsm@buaa.edu.cn}

\author{Fang Liu}
\affiliation{%
  \institution{Beihang University}
  \city{Beijing}
  \country{China}
}
\email{fangliu@buaa.edu.cn}

\author{Peiding Wang}
\affiliation{%
  \institution{Beihang University}
  \city{Beijing}
  \country{China}
}
\email{wangpeiding@buaa.edu.cn}

\author{Taichuan Li}
\affiliation{%
  \institution{Beihang University}
  \city{Beijing}
  \country{China}
}
\email{taichuanli@buaa.edu.cn}

\author{Yinghao Zhu}
\affiliation{%
  \institution{The University of Hong Kong}
  \city{Hong Kong}
  \country{China}
}
\email{yhzhu99@gmail.com}

\author{Xiaoli Lian}
\affiliation{%
  \institution{Beihang University}
  \city{Beijing}
  \country{China}
}
\email{lianxiaoli@buaa.edu.cn}

\author{Li Zhang}
\affiliation{%
  \institution{Beihang University}
  \city{Beijing}
  \country{China}
}
\email{lily@buaa.edu.cn}

\renewcommand{\shortauthors}{Liu et al.}

\begin{abstract}

As software systems grow increasingly complex, automated vulnerability repair (AVR) remains difficult because the materials available to a repair system are usually failure artifacts rather than repair guidance. Traditional analysis techniques can provide suspicious locations, reduced triggers, or constraints, but they are costly to configure across repositories and seldom directly actionable for patch generation. Recent LLM-based agents can edit and validate repository-level patches, and experience-based systems can reuse prior repair traces or demonstrations, but they still need current-instance evidence that turns a broad, symptom-level failure report into a concrete repair decision.
We present \tool{}, an agentic AVR framework that constructs such evidence through contrastive runtime analysis. Starting from a failing witness, \tool{} generates nearby failing and non-failing variants, executes them through aligned probe sites, and compares their runtime states to infer the repair boundary and guide source-level patching. Each candidate patch is accepted only after build and validation. \tool{} also stores validated repair episodes in a dual-track skill base, reusing mutation skills to construct useful variants and correction skills to refine failed patches. On SEC-Bench, \tool{} with GPT-5-mini achieves resolution rate of 92.0\% over three repeated runs and an average resolution rate of 91.8\% $\pm$ 0.8. On PatchEval, it resolves 73.8\% of 225 Go, Python, and JavaScript instances. A semantic audit of benchmark-validated SEC-Bench patches shows that 58.2\% of \tool{}'s patches are semantically correct, compared with 31.3\% for the strongest baseline, indicating that the proposed framework improves semantic correctness beyond benchmark validation. We provide the replication package at \url{https://figshare.com/s/f173c78e44bca88ebaea}

\end{abstract}


\maketitle

\section{Introduction}
\label{sec:intro}

Modern software systems are growing rapidly in scale and complexity. As real-world systems expand, security vulnerabilities emerge more frequently and become harder to manage~\cite{blogReviewZeroday, cve, yu2025patchagent}. If left unpatched, these vulnerabilities can cause data leakage, privilege escalation, service disruption, and attacks on critical infrastructure, imposing substantial economic and operational costs~\cite{anwar2020measuring, cybersecurityventuresCybercrimeCost}. Although fuzzing, static analysis, and sanitizer-based testing have automated vulnerability discovery~\cite{serebryany2017oss, manes2019fuzzingSurvey}, repair still largely relies on human experts who must reproduce the failure, understand the vulnerable path, identify the root cause, design a security check, and validate the patch. This cost does not scale with modern disclosure volume~\cite{cyware2024challenges, vicarius2024remediation}. Recent large language models (LLMs) and agentic repair systems, which can generate code, inspect repositories, and interact with execution tools, have therefore made Automated Vulnerability Repair (AVR) increasingly promising~\cite{hu2025sok,li2025sokeffectiveautomatedvulnerability,pearce2023examining,xia2023plastic}.

\begin{figure}[H]
    \centering
    \includegraphics[width=1.05\linewidth]{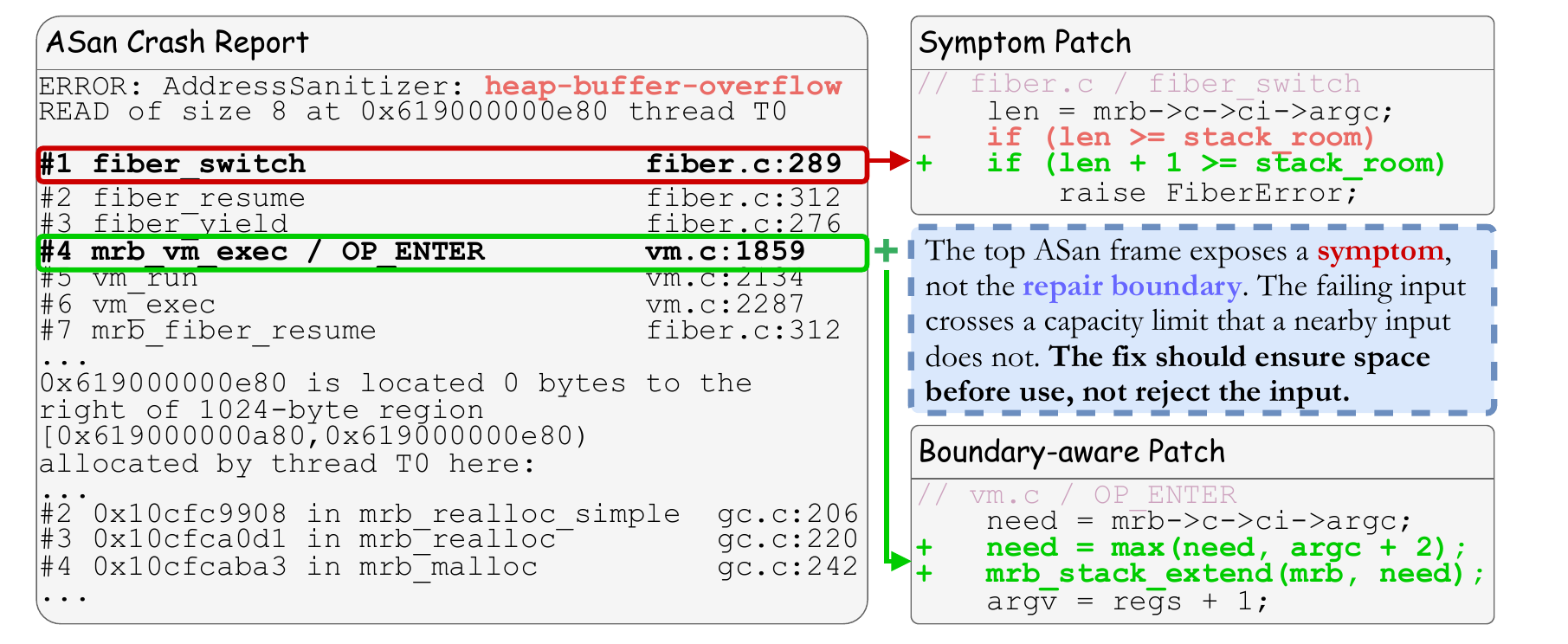}
    \caption{Motivating example of symptom-driven failure in AVR.}
    \label{fig:motivation}
\end{figure}

Before LLM-based agents, AVR mainly relied on program analysis and learning-based repair. Static analysis, path exploration, exploit-neighborhood analysis, input reduction, and constraint-based repair can produce useful suspicious locations, reduced triggers, or repair constraints~\cite{VulnLoc,zeller2002simplifying,10.5555/3489212.3489226,SemFix,CrashRepair}. However, these techniques are costly to configure across large repositories and heterogeneous build systems, and their outputs are rarely patch-ready. A patcher still has to decide the source-level boundary condition, insertion point, and repair scope. Learning-based repair systems trained on historical fixes can synthesize edits~\cite{fu2022vulrepair,jiang2021cure,chen2023vrepair}, but they typically operate on function-level snippets or pre-localized regions, and do not solve repository-level localization and validation.
LLM-based repair agents improve this situation by making repository-level repair more practical. In issue-driven APR tasks such as SWE-bench~\cite{jimenez2023swe}, agents inspect code, search repositories, generate edits, and validate candidates against tests. Recent AVR agents, including PatchAgent~\cite{yu2025patchagent}, SAN2PATCH~\cite{kim2025SAN2PATCH}, and VulnResolver~\cite{zhang2026vulnresolver}, adapt this workflow to vulnerability repair by using vulnerability reports, crash logs, retrieved fixes, security reasoning, and iterative validation. Another line further augments LLM-based repair with experience retrieval or memory \cite{Bouzenia2025RepairAgent}. ExpeRepair~\cite{ExpeRepair} reuses demonstrations, traces, and semantic insights from previous repairs, while Agenticrepair~\cite{AgentMem} accumulates repair histories or interaction traces to guide later attempts.

These systems improve patch generation, validation, and reuse, but they still depend on the quality of the available repair evidence. In AVR, the initially supplied evidence usually demonstrates an unsafe execution rather than the intended repaired behavior. A vulnerability report, PoC, sanitizer log, or failing security test can reproduce the failure \cite{PoCGen}, but it often leaves the repair under-specified. The call stack may contain many plausible project frames, and the crash site may be where invalid state is finally consumed rather than where the repair boundary should be enforced. Experience reuse does not remove this problem by itself: if retrieved traces or previous edits were guided by a symptom-level crash site or PoC-specific validation, reuse can repeat the wrong condition, location, or repair scope.
Figure~\ref{fig:motivation} illustrates why repair localization is difficult in AVR. The original repair materials include a PoC, an ASan report, and a stack trace, but these artifacts provide mixed signals rather than a direct repair decision. Several stack frames and nearby code regions look relevant, and each suggests a different possible fix, such as adding a local guard at the crash-adjacent API, rejecting the observed input pattern, or checking the upstream state that later causes the overflow. A symptom-driven agent may therefore follow the most visible ASan frame and patch the location where the invalid state is consumed. However, the correct repair depends on a boundary that is not explicit in the report: a nearby valid input still fits within the VM stack, while the failing input crosses the capacity boundary. By comparing these nearby passing and failing executions, \tool{} identifies the repair-relevant boundary and guides the patch toward ensuring enough stack space before use, rather than merely rejecting the input.

Motivated by the gap, we propose \tool{}, an agentic AVR framework that turns a single exploit witness into boundary evidence for localization, patch generation, and experience reuse.
\noindent 1) \textbf{Contrastive PoC localization.} \tool{} constructs format-preserving \textsc{same-crash} and \textsc{oracle-passing} PoC variants that still reach the same vulnerable execution region, then compares their executions to narrow a broad crash context to the failure boundary.
\noindent 2) \textbf{Patch-ready boundary evidence.} \tool{} records concrete runtime values at shared probe sites and summarizes the boundary evidence into repair guidance, including where to patch, what condition to enforce, and what scope the patch should cover.
\noindent 3) \textbf{Reusable AVR skills.} \tool{} accumulates two kinds of validated experience. Mutation skills help future runs construct useful contrastive PoCs, while correction skills record how a failed patch was later refined into a validated repair.

We evaluate \tool{} on SEC-Bench~\cite{lee2025secbench}, which contains 200 C/C++ vulnerability instances, and PatchEval~\cite{wei2025patcheval}, which contains 225 usable Go, Python, and JavaScript instances. With GPT-5-mini as the backbone, \tool{} achieves state-of-the-art performance on both benchmarks, reaching 92.0\% resolution rate on SEC-Bench and resolving 73.8\% of PatchEval. On SEC-Bench, \tool{} outperforms Agenticrepair \cite{AgentMem} by 42.0 percentage points
under the same GPT-5-mini backbone, supporting the conclusion that the framework contributes substantially beyond backbone capability alone.
Ablation results show that contrastive runtime analysis provides a 26.8-point average gain, while the full skill-accumulation setting adds a further 17.0 points over contrastive-only analysis.
In summary, this paper makes the following contributions:
\begin{itemize}[leftmargin=*]

\item \textbf{Boundary-Oriented PoC Analysis for AVR.}
We introduce a runtime analysis method that compares aligned failure-reproducing and oracle-passing executions, identifies runtime relations that distinguish the observed execution classes, and encodes those relations and their supporting evidence in location-specific repair specifications.

\item \textbf{Vulnerability-Specific Skill Accumulation.}
We design a skill mechanism tailored to AVR that accumulates boundary-construction skills for PoC mutation and failure-to-success refinement skills for revising benchmark-invalid candidate patches into benchmark-validated repairs.

\item \textbf{Extensive Evaluation.}
We evaluate \tool{} on SEC-Bench and PatchEval across C/C++, Go, Python, and JavaScript. Results show that \tool{} outperforms state-of-the-art AVR baselines and general-purpose software engineering agents, with ablations confirming the complementary benefits of contrastive analysis and skill accumulation.

\end{itemize}

\section{Proposed Framework}
\label{sec:method}

\begin{figure*}[t]
    \centering
    \setlength{\abovecaptionskip}{0.1cm}
     \includegraphics[width=0.87\textwidth]{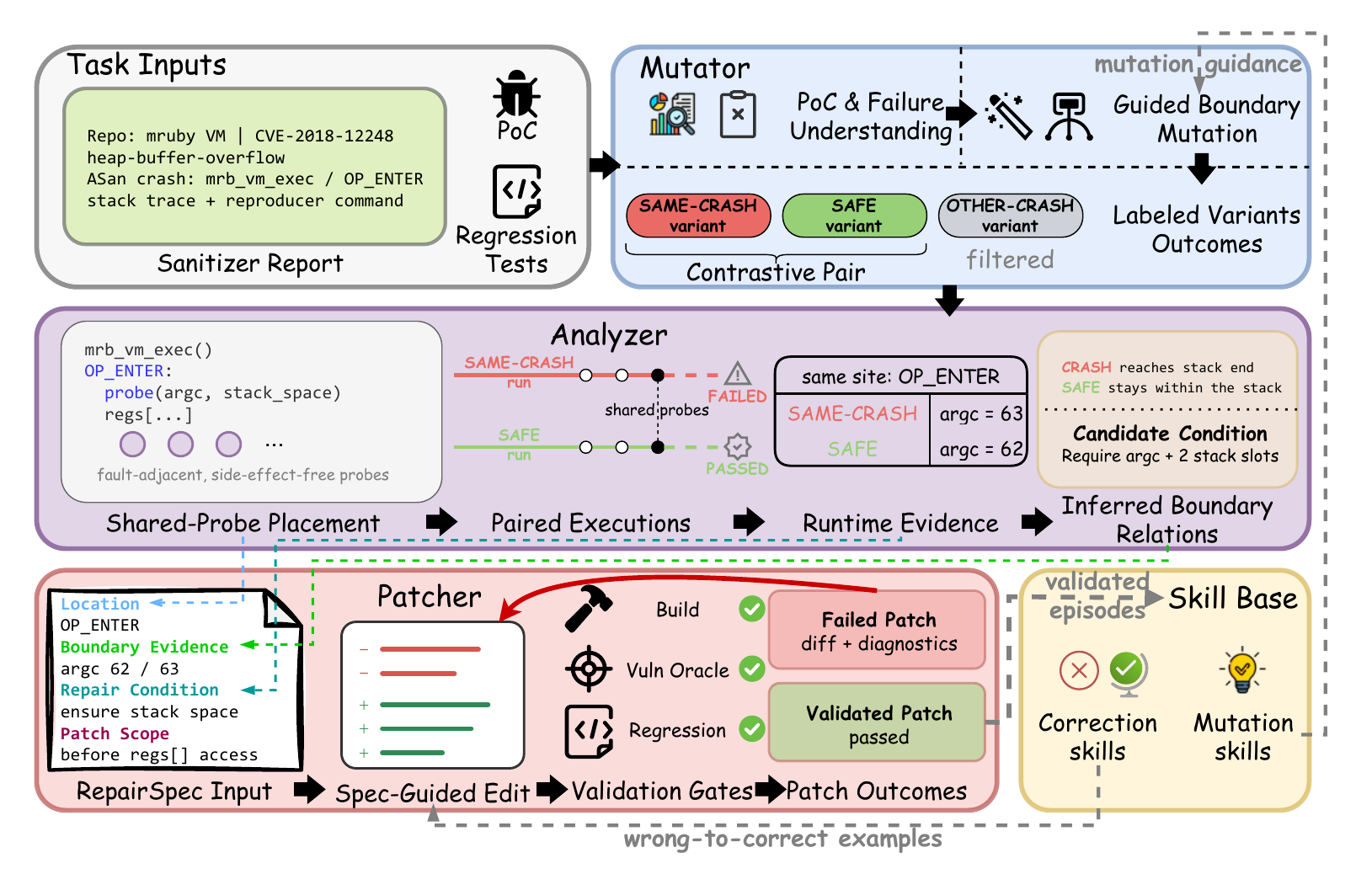}
    \caption{Overall architecture of \tool{}.}
    \label{fig:overview}
    \vspace{-0.3cm}
\end{figure*}

Figure~\ref{fig:overview} illustrates the overall pipeline of \tool{}. Three coordinated agents operate in an iterative repair loop. The \textbf{Mutator} constructs PoC variants around the observed failure boundary (Section~\ref{sec:mutator}). The \textbf{Analyzer} compares failure-reproducing and oracle-passing executions and formulates a location-specific RepairSpec (Section~\ref{sec:analyzer}). The \textbf{Patcher} translates the RepairSpec into a candidate source patch (Section~\ref{sec:patcher}). Benchmark-validated episodes are recorded in a dual-track skill base for subsequent instances (Section~\ref{sec:skill}).

The repair loop is gated by the results of concrete runs. Each PoC variant is executed and labeled by its outcome. The Analyzer summarizes observed state differences at shared code locations into repair guidance. The Patcher accepts a candidate patch only if it builds, no longer triggers the vulnerability on the original PoC and same-crash variants, and passes the benchmark regression tests. When a patch fails, its diff, diagnostics, and prior analysis context are carried into the next round. Table~\ref{tab:tools} summarizes the role permissions, while the complete prompts and machine-readable schemas are included in the replication package.

\subsection{PoC Variant Generation}
\label{sec:mutator}
The Mutator generates format-valid variants that make small, targeted changes to the original PoC and executes each variant to observe its outcome. A variant is useful for contrastive analysis when it either reproduces the original failure signature or no longer triggers the vulnerability oracle while still reaching the relevant code. When a failure-reproducing execution and an oracle-passing execution reach the same probe site, their recorded runtime states form contrastive evidence that the Analyzer can use to infer candidate state relations for repair.

\paragraph{Vulnerability-Aware Mutation Strategies}
\label{sec:mutation-strategy}

Real-world PoCs span diverse formats, from structured binary containers (e.g., MP4, TIFF, DWG) to interpreter scripts (e.g., Ruby, JavaScript, PHP). Effective mutation must respect the structural constraints of each format; blindly corrupting bytes typically produces unparseable inputs rejected long before execution reaches the vulnerable code path~\cite{zalewski2014afl, fioraldi2020aflplusplus}. \tool{} addresses this challenge with a two-level mutation strategy.

At the \emph{format level}, the Mutator examines the original PoC and its file signature to select an appropriate manipulation method. For binary containers it generates a Python script that operates on semantic fields (e.g., box sizes and sample counts in MP4, dimension tags in TIFF, chunk lengths in DWG) while preserving magic bytes, checksums, and framing headers. For interpreter scripts it rewrites the source directly, modifying literal values, loop bounds, or argument lists that control the execution path reaching the fault site.

At the \emph{vulnerability-type level}, we initialize the mutation skill base with author-defined mutation templates derived from the sanitizer failure classes observed in SEC-Bench~\cite{lee2025secbench}. These templates are manually designed from the failure semantics of each class and serve only as cold-start mutation guidance; the concrete mutation is still generated after the Mutator inspects the current PoC format, execution command, and failure report. For example, overflow-related templates encourage changes to size, count, offset, and index fields, while lifetime- and nullness-related templates guide mutations around allocation, release, reuse, and missing-reference patterns. The complete template list is included in the code repository. As instances are resolved, the skill base accumulates format-specific strategies from successful resolutions, progressively supplementing the generic templates with empirically validated ones.

\paragraph{Differential Variant Generation}
The goal of this step is to obtain two executions that differ only in whether the vulnerability fires: one that still triggers the \emph{same} bug as the original PoC, and one that runs safely past the same code. Comparing this pair later tells the Analyzer which runtime value actually decides the outcome.

The Mutator starts from the original PoC $x_0$ and produces candidate variants with format-aware and vulnerability-aware transformations, running each through \texttt{run\_variant}. For every variant $x$ we record three things: whether it triggers the vulnerability ($O(x)\!=\!1$ if it does), its \emph{failure signature} $\mathrm{sig}(x)$ (for sanitizer cases, the sanitizer class plus a project-level stack frame; for test cases, the failing test plus its error), and the set of probe sites it actually reaches, $\Pi(x)$. Each variant is then sorted into one of three groups:
\begin{equation}
\mathrm{label}(x)=
\begin{cases}
\textsc{same-crash}, & \text{triggers the same
bug} \, x_0,\\
\textsc{other-crash}, & \text{triggers a different failure},\\
\textsc{oracle-passing}, & \text{does not trigger the bug}.
\end{cases}
\end{equation}
A useful comparison needs one variant from the first group and one from the third. But a safe-looking variant is only meaningful if it actually runs through the same place as the crash—otherwise it might look safe simply because parsing stopped early, long before the vulnerable code. We therefore require the safe variant $x^+$ and the crashing variant $x^-$ to share at least one probe site that the original failing run also passed through:
\begin{equation}
\Pi(x^-)\cap\Pi(x^+)\cap\Pi(x_0)\neq\varnothing .
\end{equation}
We call such a pair an \emph{ideal contrastive pair}.

\begin{table}[t]
\centering
\footnotesize
\setlength{\tabcolsep}{2.5pt}
\renewcommand{\arraystretch}{1.05}
\caption{Role-restricted tool access in \tool{}. M/A/P denote Mutator,
Analyzer, and Patcher, respectively. Complete tool schemas and prompts are
provided in the replication package.}
\label{tab:tools}
\begin{tabularx}{\linewidth}{@{}L{0.25\linewidth}Yccc@{}}
\toprule
\textbf{Toolkit} & \textbf{Representative tools} & \textbf{M} & \textbf{A} & \textbf{P} \\
\midrule
File inspection & \texttt{view}, \texttt{search} &  & \cmark & \cmark \\
File editing & \texttt{str\_replace\_edit}, \texttt{revert\_*} &  & \cmark & \cmark \\
Execution & \texttt{bash} & \cmark & \cmark & \cmark \\
PoC mutation & \texttt{mutate\_poc}, \texttt{run\_variant} & \cmark &  &  \\
State probing & \texttt{insert\_probed}, \texttt{run\_variant}  &  & \cmark &  \\
Validation & \texttt{check\_vul} &  &  & \cmark \\
\bottomrule
\end{tabularx}
\vspace{-0.2cm}
\end{table}

\subsection{Contrastive Analysis}
\label{sec:analyzer}

The Analyzer converts evidence from an aligned execution pair into a source-level RepairSpec. Given a pair in which one execution reproduces the original failure, the other produces an oracle-passing outcome, and both reach a shared probe site, the Analyzer localizes the fault region, probes candidate runtime expressions, and summarizes relations that distinguish the observations collected from the two executions. The goal is not to establish causality or prove a program invariant, but to derive a concrete, location-specific candidate condition that can guide patch generation.

\paragraph{Phase 1: Fault Location}
The Analyzer extracts the failure type, project-level stack frames, and faulting source region from the sanitizer or test report. Candidate probe expressions are drawn from three concrete sources: operands of the faulting statement, variables used by nearby controlling predicates, and security-relevant fields referenced in the fault function or its inspected
callers. These fields include indices, lengths, capacities, allocation sizes, pointer/nullness states, ownership flags, and parser-state variables. The Analyzer probes only side-effect-free expressions that can be evaluated at a specific source location.

\paragraph{Phase 2: Contrastive State Probing}
The Analyzer instruments the candidate fault region with lightweight language-level probes and executes the same instrumented program on both same-crash and oracle-passing variants. \tool{} supports two probe forms: a \emph{point probe}, which records a snapshot immediately before or after a suspicious statement, and a \emph{span probe}, which records entry/exit snapshots around a short code region. Each probe record contains the probe site, source location, execution phase,
and concrete values of the selected candidate expressions. The complete probe schema and language-specific logging templates are provided in the replication package.

Probe records are keyed by a stable probe identifier, source location, execution phase, and variant identifier. The Analyzer compares only records observed at a common probe site; unmatched records are excluded rather than treated as evidence. It examines simple security-relevant relations, including index--capacity and size--allocation bounds, nullness, equality/inequality
conditions, and state transitions. A relation is retained when the available oracle-passing observations satisfy it while a same-crash observation violates it.

For example, if a probe before a buffer access records
\texttt{idx=15, cap=16} in an oracle-passing execution and
\texttt{idx=16, cap=16} in an execution that reproduces the original failure, then \texttt{idx < cap} distinguishes the two observations. This relation is supported by the sampled executions at that probe site, but we do not claim that it establishes causality or constitutes a formally proven program invariant.

\paragraph{Phase 3: RepairSpec Inference}
The Analyzer converts the retained evidence into a fixed-format RepairSpec:
\textbf{Location} identifies the file, function, and source range;
\textbf{Boundary Evidence} cites the probe sites and concrete values observed in the failure-reproducing and oracle-passing executions;
\textbf{Candidate Repair Condition} records the relation proposed as repair guidance; and \textbf{Patch Scope} identifies the state-creation, validation, or use sites at which the candidate condition may need to be enforced. The prompt requires every proposed condition to be supported by recorded probe values. When the shared-site evidence does not yield a clear separating relation, the Analyzer summarizes the available observations and uncertainty in the RepairSpec. The RepairSpec is treated as repair guidance rather than a proof, and the generated patch must still pass the normal build, vulnerability-oracle, and regression-oracle gates. The exact prompt and schema are included in the replication package.

Returning to the motivating example in Figure~\ref{fig:motivation}, for CVE-2018-12248~\cite{cve-2018-12248}, probe records from the
\texttt{OP\_ENTER} handler show \texttt{argc=62} in the oracle-passing execution and \texttt{argc=63} in the failure-reproducing execution. The execution with \texttt{argc=62} remains within the allocated fiber stack, whereas the execution with \texttt{argc=63} reaches the stack end \texttt{stend}. Based on these observations, the RepairSpec proposes ensuring that the VM stack has space for at least \texttt{argc+2} slots before the first \texttt{regs[]} access in \texttt{src/vm.c:mrb\_vm\_exec}.

\paragraph{Cross-Round Refinement}
When a candidate patch fails PoC validation, \tool{} starts a new round with three pieces of carry-over evidence: the failed diff and diagnostic output, the previous RepairSpec, and the prior probe plan. This allows the Analyzer to refine the previous hypothesis rather than repeat the same analysis from
scratch. After each analysis phase, all probe edits are reverted to ensure that the final submitted diff contains only the Patcher's source changes.

\subsection{Patch Generation}
\label{sec:patcher}

The Patcher translates the RepairSpec into a candidate source patch, using retrieved benchmark-validated trajectories as repair examples. The examples provide corrective patterns rather than acceptance evidence: every transferred edit is regenerated for the current repository and independently validated.

All source modifications are applied through str\_replace\_edit, and the controller extracts the resulting diff from the working tree. A candidate is counted as benchmark-validated only if it passes three gates:
(1) the project builds successfully; (2) the benchmark vulnerability oracle passes on the original PoC and all available same-crash variants; and (3) the benchmark regression oracle passes. Failure at any gate supplies the next round with the exact diff and the corresponding compiler, sanitizer, or
test diagnostics. Probe instrumentation is reverted before patch validation, so the submitted diff contains only Patcher edits.

After each round, \texttt{revert\_all\_edits()} restores a clean working tree before the next candidate is generated.

\subsection{Experience Accumulation}
\label{sec:skill}

\tool{} records two experience tracks across resolved instances. The first track supports PoC mutation. A mutation record stores
$\langle P_{repo}, K_{vuln}, D_{san}, F_{poc}, S_{mut}, V\rangle$, where $F_{poc}$ describes the input family, $S_{mut}$ summarizes the effective transformation, and $V$ stores per-variant outcomes such as mutation operation, oracle result, sanitizer class, and a short output snippet. These records let the Mutator reuse format-preserving transformations, such as changing MP4 box sizes, TIFF count/offset fields, or script literals while keeping the input on the vulnerable path.

The second track supports patch correction. A correction record is created only when an instance is validated after at least one failed patch. It stores the repository, vulnerability class, failed diffs with compiler/oracle feedback, and the final benchmark-validated diff. During a later repair, the Patcher retrieves these records only after its current candidate fails, and uses them as wrong-to-correct demonstrations, such as moving a symptom-side guard upstream or applying the same bounds check to sibling paths.

The two tracks are retrieved independently with a default budget of two entries each. Retrieval first prefers the same repository and vulnerability type, then falls back to same-type cross-repository episodes. To prevent leakage, \tool{} excludes same-CVE and later-disclosed entries, and inserts the current instance into the skill base only after its repair run terminates. Retrieved experience guides generation but never bypasses build, vulnerability-oracle, or regression validation.

\section{Experimental Setup}
\label{sec:setup}

\subsection{Research Questions}

We evaluate \tool{} with the following research questions:
\begin{itemize}[leftmargin=*]
\item \textbf{RQ1: Overall Performance, Localization Accuracy, and Patch Correctness.}
How does \tool{} compare with existing repair methods in benchmark validation and localization accuracy, and how many of its benchmark-validated patches are semantically correct?

    \item \textbf{RQ2: Ablation Study.} How much do contrastive state probing and experience accumulation contribute to the final repair rate?

    \item \textbf{RQ3: Generalizability.} Does \tool{} generalize from C/C++ to Go, Python, and JavaScript vulnerabilities?

\end{itemize}

\subsection{Benchmarks and Metrics}

We evaluate on two execution-based vulnerability repair benchmarks.

\textbf{SEC-Bench}~\cite{lee2025secbench} contains 200 real-world CVE instances from 29 C/C++ projects. Each instance provides a vulnerable repository, a PoC, reproduction commands, and sanitizer feedback. We use SEC-Bench as the primary benchmark because it evaluates end-to-end repository-level AVR under sanitizer-based vulnerability oracles.

\textbf{PatchEval}~\cite{wei2025patcheval} covers Go, Python, and JavaScript vulnerabilities. We use 225 Dockerized instances that can be built and executed in our environment; five instances are excluded before evaluation because their Docker images cannot be pulled. We use PatchEval to test whether \tool{} generalizes beyond C/C++ memory-safety settings.

We report \textbf{\%Resolved}, \textbf{FileLoc}, \textbf{HunkLoc}, and \textbf{Avg.\ Cost}. A patch is counted as resolved only if it passes the benchmark's execution-based vulnerability and regression oracles. We evaluate localization following the rule~\cite{yang2024sweAgent}: FileLoc counts an instance as correctly localized if the generated patch edits all developer-modified files, and HunkLoc counts an instance as correctly localized if generated hunks overlap all developer-modified hunks. Avg.\ Cost is the average LLM API cost per instance in USD, computed from token usage. To distinguish benchmark validation from semantic correctness, we additionally audit benchmark-validated patches using three categories: semantically correct, incomplete/PoC-specific, and behavior-regressive.
We use same-backbone GPT-5-mini comparisons for framework attribution and report cross-backbone comparisons separately.

\subsection{Baselines and Adaptation}

We compare \tool{} with four baseline groups. General-purpose coding agents include OpenHands~\cite{wang2024openhands}, smolagents~\cite{smolagents}, SWE-agent~\cite{yang2024sweAgent}, and Claude Code~\cite{claudecode2025}. AVR-specific agents include PatchAgent~\cite{yu2025patchagent}, VulnResolver~\cite{zhang2026vulnresolver}, and Agenticrepair~\cite{AgentMem}. Experience-based repair is represented by ExpeRepair~\cite{ExpeRepair}. Constraint-guided repair is represented by CrashRepair+LLM, where the LLM receives the vulnerable-function context and a CrashRepair-style repair constraint~\cite{CrashRepair} before synthesizing a source patch.

We preserve each baseline's native repair workflow, memory, and hyperparameters unless the baseline is explicitly adapted, as in CrashRepair+LLM. For ExpeRepair, we keep the original implementation and replace only the execution wrapper so its patches are checked by the same SEC-Bench oracle as \tool{}. SEC-Bench patches are evaluated with sanitizer-based workflows; PatchEval patches are evaluated with the benchmark's language-specific tests, including \texttt{go test}, \texttt{pytest}, and \texttt{npm test}. We exclude older function-level AVR systems evaluated mainly with static similarity metrics such as CodeBLEU~\cite{codebleu}, because they do not operate in the same repository-level, execution-validated setting.
\subsection{Implementation Details}

We use GPT-5-mini as the default backbone for all three \tool{} agents and rerun LLM-based baselines unless otherwise stated. On SEC-Bench, we run the full \tool{} configuration three times, report the median run by resolved instances, and use that run for per-instance analyses. The repair loop runs for at most three rounds; a sensitivity run extends the budget to five rounds. All LLM calls use the model default decoding setting, which is temperature 1.0 and top-p 1.0 for GPT-5-mini.

All experiments run inside isolated Docker containers provisioned from the benchmark images. For SEC-Bench, \texttt{run\_variant} and \texttt{check\_vul} wrap the benchmark's sanitizer workflow and use AddressSanitizer signals to classify executions. For PatchEval (Go, Python, JavaScript), which is test-driven, these tools instead invoke the respective language's testing frameworks (\textit{e.g.}, \texttt{go test}, \texttt{pytest}, \texttt{npm test}).

\section{Results and Analysis}\label{sec:results}

\subsection{RQ1: Overall Performance, Localization Accuracy, and Patch Correctness}\label{sec:rq1}

RQ1 evaluates \tool{} from four perspectives: benchmark validation, localization accuracy, semantic correctness, and performance across vulnerability types. The RQ1 tables report the median SEC-Bench run for \tool{}, which also supplies the patches used in the semantic and per-instance analyses.

\paragraph{Overall benchmark validation and localization accuracy}
Table~\ref{tab:rq1-overall} compares \tool{} with general-purpose coding agents, AVR-specific agents, experience-based repair, and constraint-guided repair on SEC-Bench. In the median SEC-Bench run, \tool{} with GPT-5-mini resolves 184/200 instances (92.0\%), achieving the highest benchmark-validation rate among all evaluated configurations. The strongest baseline is Agenticrepair with GPT-5.2, which resolves 150/200 instances (75.0\%). \tool{} improves over it by 17.0 percentage points and repairs 34 additional instances. Under the same GPT-5-mini backbone, \tool{} outperforms PatchAgent, ExpeRepair, Agenticrepair, OpenHands, and Smolagents by 29.5, 32.0, 42.0, 43.0, and 57.5 points, respectively. The best general-purpose baseline, OpenHands, reaches 49.0\%, suggesting that repository-level coding ability alone is insufficient for AVR.

The comparison also helps separate \tool{}'s framework contribution from two simpler alternatives. First, CrashRepair+LLM receives vulnerable-function context and a CrashRepair-style constraint, but resolves only 29.0\% of instances. This suggests that a repair constraint alone still leaves important source-level decisions unresolved, including where to enforce the condition, how far the check should be propagated, and how to preserve normal behavior. Second, \tool{} remains strong under a different backbone: with DeepSeek-V3.2~\cite{liu2025deepseek}, it resolves 178/200 instances (89.0\%), still outperforming all evaluated baselines. As shown in the cost column, \tool{} with DS-V3.2 averages \$0.15 per task, lower than PatchAgent-GPT-5-mini (\$0.40) and less than one-third of Agenticrepair-GPT-5.2 (\$1.07). With the default GPT-5-mini backbone, \tool{} costs \$0.32 per task while achieving the highest resolution rate.

For localization accuracy, \tool{} with GPT-5-mini achieves 80.5\% FileLoc and 59.0\% HunkLoc, close to the highest localization scores while achieving substantially higher benchmark validation. Agenticrepair-GPT-5.2 obtains the highest FileLoc and HunkLoc, 82.5\% and 62.5\%, respectively. This result is partly due to patch breadth. Its generated patches contain on average 2.2$\times$ as many hunks as \tool{}'s patches, which increases the chance that the generated file and hunk sets cover the developer-modified locations used by FileLoc and HunkLoc. Its multi-agent analysis also provides static-analysis results, historical commits, and multiple repair suggestions, so the final patch often combines several plausible repair ideas and edits a broader region. The manual audit below further shows that higher FileLoc and HunkLoc do not necessarily imply semantic correctness.

\begin{table}[t]
  \centering
  \footnotesize
  \setlength{\abovecaptionskip}{0.1cm}
  \caption{Overall performance, localization accuracy, and cost on SEC-Bench.}
  \label{tab:rq1-overall}
  \setlength{\tabcolsep}{2.0pt}
  \renewcommand{\arraystretch}{1.06}
  \resizebox{\linewidth}{!}{%
  \begin{tabular}{@{}l c c c c c@{}}
    \toprule
    \textbf{Method} & \textbf{Backbone} & \textbf{\%Res.} & \textbf{FileLoc} & \textbf{HunkLoc} & \textbf{Avg.\ Cost} \\
    \midrule
    OpenHands~\cite{wang2024openhands}        & GPT-5-mini    & 49.0 & 58.0 & 51.5 & 0.22 \\
    Smolagents~\cite{smolagents}              & GPT-5-mini    & 34.5 & 60.0 & 52.0 & 0.04 \\
    Smolagents~\cite{smolagents}              & GPT-5.2       & 45.0 & 67.0 & 58.5 & 0.50 \\
    \midrule
    CrashRepair+LLM~\cite{CrashRepair}        & GPT-5-mini    & 29.0 & --   & --   & 0.01 \\
    Agenticrepair~\cite{AgentMem}             & GPT-5-mini    & 50.0 & 67.5 & 57.5 & 0.16 \\
    ExpeRepair~\cite{ExpeRepair}              & GPT-5-mini    & 60.0 & 63.0 & 45.0 & 0.22 \\
    PatchAgent~\cite{yu2025patchagent}        & GPT-5-mini    & 62.5 & 58.5 & 42.5 & 0.40 \\
    VulnResolver~\cite{zhang2026vulnresolver} & DS-V3.2-Exp   & 67.5 & --   & --   & 0.07 \\
    Agenticrepair~\cite{AgentMem}             & GPT-5.2       & 75.0 & \textbf{82.5} & \textbf{62.5} & 1.07 \\
    \midrule
    \tool{}                                   & GPT-5-mini    & \textbf{92.0} & 80.5 & 59.0 & 0.32 \\
    \tool{}                                   & DS-V3.2       & \underline{89.0} & 80.5 & 58.5 & 0.15 \\
    \bottomrule
  \end{tabular}%
  }
  \vspace{-0.25cm}
\end{table}

\begin{table*}[t] \centering \tabledoublesetup \setlength{\abovecaptionskip}{0.1cm} \caption{Resolution rate (\%) by vulnerability type on SEC-Bench under GPT-5-mini. Values in parentheses denote resolved instances.} \label{tab:rq1-vulntype} \setlength{\tabcolsep}{3pt} \renewcommand{\arraystretch}{1.12} \begin{tabularx}{\textwidth}{@{}L{0.22\textwidth} C{0.05\textwidth} *{5}{Z}@{}} \toprule \textbf{Vulnerability Type} & \textbf{$N$} & \textbf{Smolagents} & \textbf{Agenticrepair} & \textbf{ExpeRepair} & \textbf{PatchAgent} & \textbf{\tool{}} \\ \midrule heap-buffer-overflow & 77 & \pcc{pur1}{42.9 (33)} & \pcc{pur2}{55.8 (43)} & \pcc{pur2}{59.7 (46)} & \pcc{pur3}{68.8 (53)} & \pcc{pur5}{\textbf{97.4 (75)}} \\ null-pointer-deref & 53 & \pcc{pur1}{34.0 (18)} & \pcc{pur2}{50.9 (27)} & \pcc{pur2}{58.5 (31)} & \pcc{pur2}{58.5 (31)} & \pcc{pur5}{\textbf{92.5 (49)}} \\ SEGV & 23 & \pcc{pur1}{30.4 (7)} & \pcc{pur2}{52.2 (12)} & \pcc{pur2}{56.5 (13)} & \pcc{pur2}{52.2 (12)} & \pcc{pur5}{\textbf{87.0 (20)}} \\ use-after-free & 20 & \pcc{pur1}{5.0 (1)} & \pcc{pur1}{20.0 (4)} & \pcc{pur2}{55.0 (11)} & \pcc{pur2}{60.0 (12)} & \pcc{pur4}{\textbf{80.0 (16)}} \\ memory-leak & 13 & \pcc{pur1}{23.1 (3)} & \pcc{pur1}{46.2 (6)} & \pcc{pur3}{69.2 (9)} & \pcc{pur2}{53.8 (7)} & \pcc{pur4}{\textbf{84.6 (11)}} \\ stack-buffer-overflow & 14 & \pcc{pur1}{50.0 (7)} & \pcc{pur2}{57.1 (8)} & \pcc{pur3}{71.4 (10)} & \pcc{pur3}{71.4 (10)} & \pcc{pur5}{\textbf{92.9 (13)}} \\ \midrule \textit{Overall} & 200 & \pcc{pur1}{34.5 (69)} & \pcc{pur1}{50.0 (100)} & \pcc{pur2}{60.0 (120)} & \pcc{pur2}{62.5 (125)} & \pcc{pur5}{\textbf{92.0 (184)}} \\ \bottomrule \end{tabularx} \vspace{-0.3cm} \end{table*}

\paragraph{Manual semantic review}\label{par:manual-semantic-review}
Benchmark validation does not necessarily imply semantic correctness \cite{bui2024apr4vul}, since a patch may block the supplied PoC while leaving the root cause reachable through other inputs or changing valid behavior. We therefore audited all benchmark-validated patches from \tool{}'s run and the strongest baseline configuration, covering 184 patches from \tool{} and 150 patches from Agenticrepair-GPT-5.2.
The audit classified patches into three categories: (1) semantically correct, (2) incomplete or PoC-specific, and (3) behavior-regressive.
A semantically correct patch fixes the root cause without known valid-behavior loss, including both developer-patch-aligned repairs and correct alternative repairs. Incomplete or PoC-specific patches pass the benchmark checks because they block the supplied PoC, but they do not enforce the same rule across related code paths. Behavior-regressive patches remove the trigger by rejecting or changing valid behavior. Two PhD software engineering annotators with over seven years of relevant experience independently labeled every patch. For each generated diff, they considered the vulnerability report, PoC, relevant code context, benchmark tests, and developer patch. The developer patch served as evidence of root-cause intent and repair scope rather than a syntactic oracle. Patches were shuffled and anonymized with respect to system provenance. Initial agreement was 84.8\% with Cohen's $\kappa = 0.78$, and remaining disagreements were resolved through discussion.

\begin{figure}[h]
  \centering
  \includegraphics[width=0.9\linewidth]{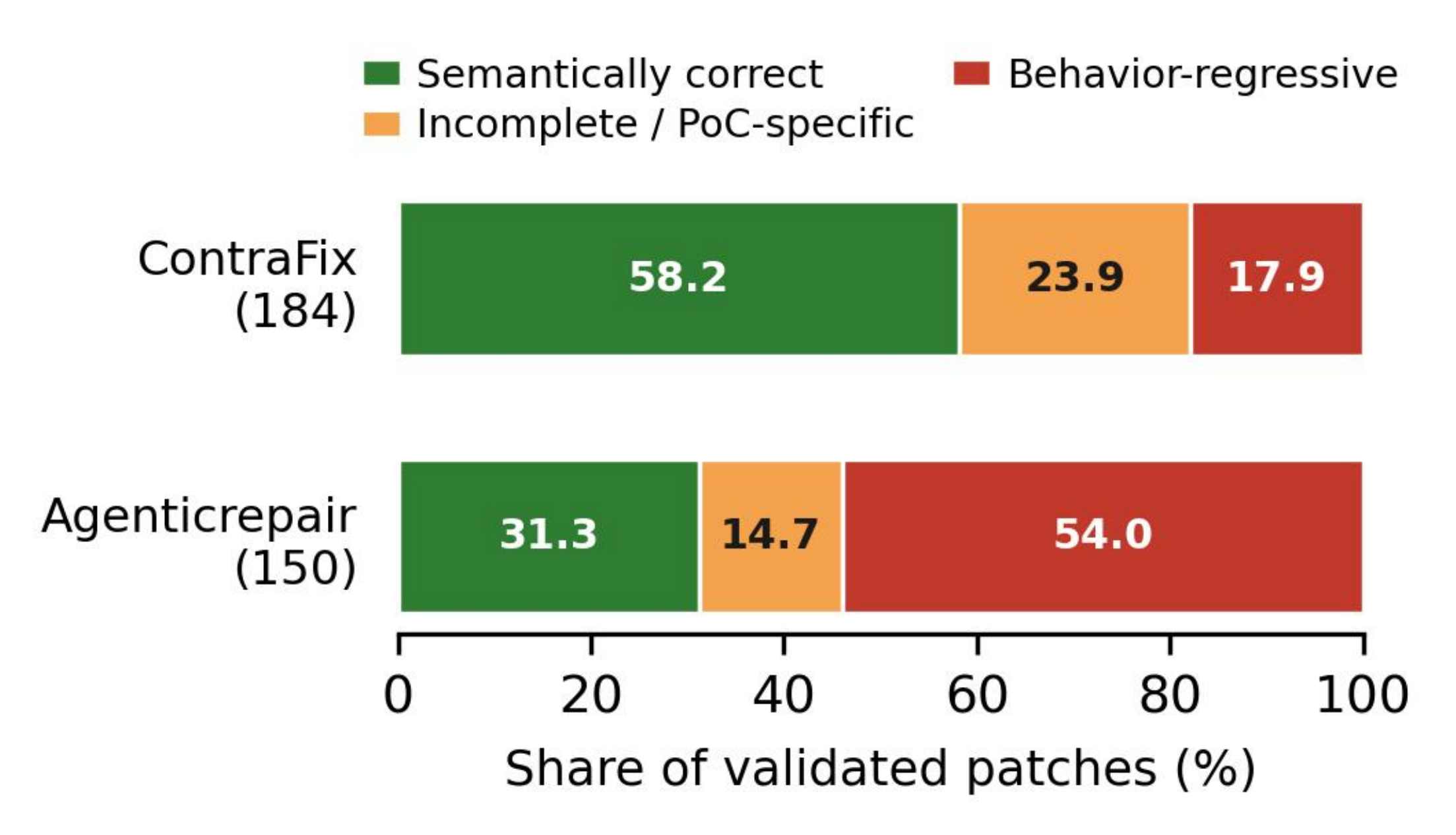}
  \caption{Manual semantic review of benchmark-validated SEC-Bench patches.}
  \label{fig:semantic-review}
  \vspace{-0.25cm}
\end{figure}

Figure~\ref{fig:semantic-review} summarizes the three-category audit. Among benchmark-validated patches, \tool{} has 107/184 semantically correct patches (58.2\%), 44 incomplete or PoC-specific patches (23.9\%), and 33 behavior-regressive patches (17.9\%), whereas Agenticrepair-GPT-5.2 has 47/150 (31.3\%), 22/150 (14.7\%), and 81/150 (54.0\%), respectively. Over all 200 SEC-Bench instances, the semantic repair rates are 53.5\% for \tool{} and 23.5\% for Agenticrepair-GPT-5.2. Together with Table~\ref{tab:rq1-overall}, these results show that higher FileLoc and HunkLoc do not necessarily imply semantic correctness. Agenticrepair-GPT-5.2 achieves the highest localization accuracy, but many of its validated patches are behavior-regressive, indicating that broad patches may cover developer-modified locations while enforcing the wrong boundary, missing the full repair scope, or rejecting valid behavior. Section~\ref{sec:failure-modes} gives concrete examples of incomplete and behavior-regressive patches.

\paragraph{Breakdown by vulnerability type}
Table~\ref{tab:rq1-vulntype} reports resolution rates by sanitizer category under GPT-5-mini. \tool{} leads in all six vulnerability types. The largest gains over PatchAgent appear on heap-buffer-overflow and null-pointer-deref, where \tool{} improves by 28.6 and 34.0 points, respectively. These defects often expose concrete bound or nullness conditions, which are well suited to same-crash/oracle-passing PoC comparison and contrastive state probing. \tool{} also remains effective on harder classes, resolving 87.0\% of SEGV and 80.0\% of use-after-free instances, where the repair condition often depends on parser state, object lifetime, or dispatch logic rather than a single local bound.

\vspace{3mm}
\begin{custommdframed}
\textbf{Answer to RQ1.} \tool{} achieves the best SEC-Bench performance, resolving 184/200 instances (92.0\%) and leading all vulnerability types, while also reaching competitive localization accuracy with 80.5\% FileLoc and 59.0\% HunkLoc. Compared with Agenticrepair-GPT-5.2, \tool{} produces far more semantically correct validated patches (58.2\% vs. 31.3\%), showing that its localization evidence better supports correct vulnerability repair.
\end{custommdframed}

\subsection{RQ2: Ablation Study}
\label{sec:rq2}

\begin{figure}[t]
  \centering
  \includegraphics[width=\linewidth]{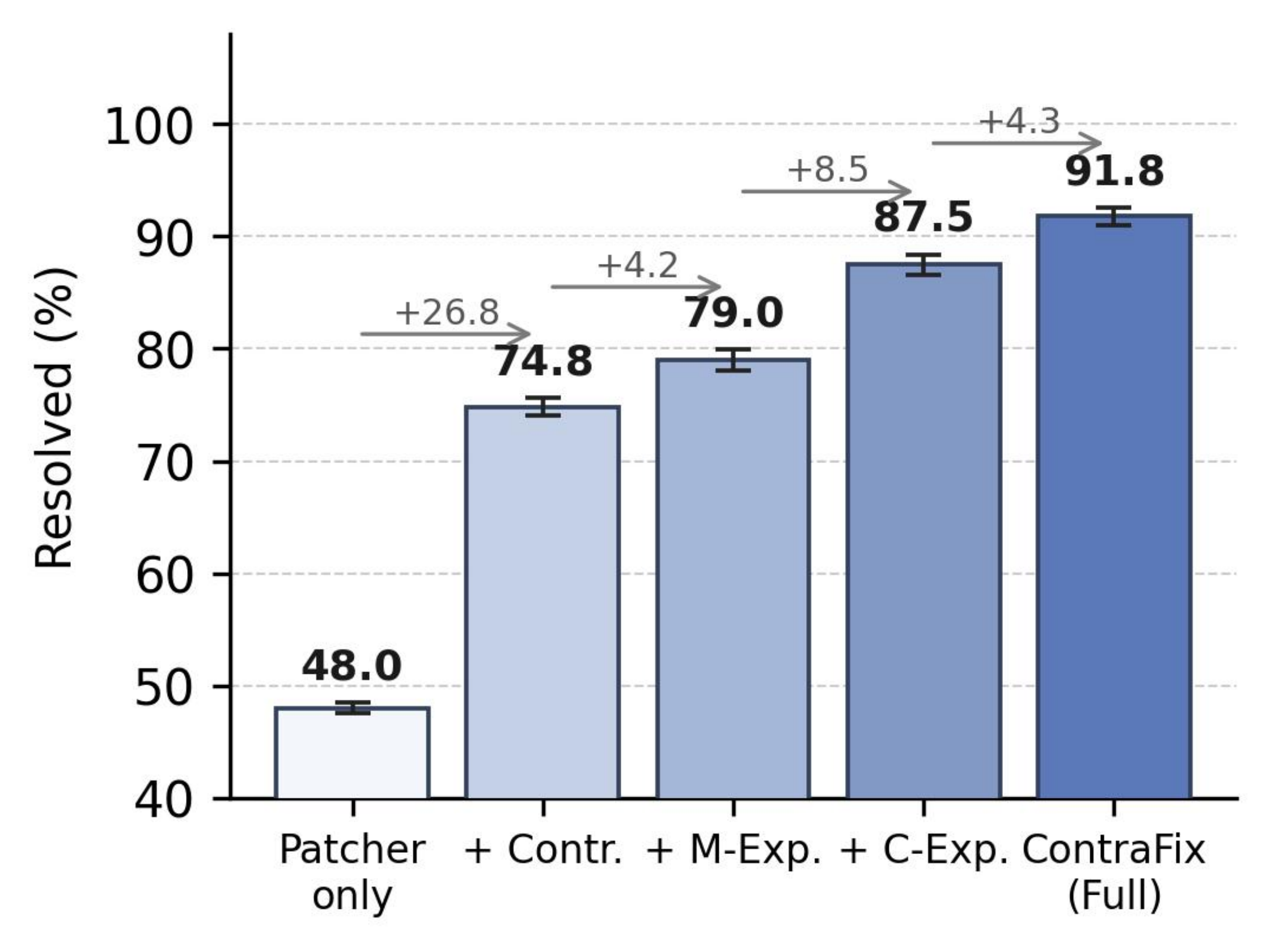}
  \caption{Ablation study on SEC-Bench over three runs.}
  \label{fig:rq2-ablation}
  \vspace{-0.25cm}
\end{figure}

Figure~\ref{fig:rq2-ablation} reports the ablation results on SEC-Bench.
We compare five configurations that progressively enable the three main components of \tool{}. \textit{Patcher only} generates a patch from the original failure report without contrastive analysis or experience retrieval. \textit{+ Contr.} enables the Mutator and Analyzer, so the Patcher receives a RepairSpec inferred from contrastive runtime evidence. \textit{+ M-Exp.} additionally enables mutation experience to guide PoC variant construction. \textit{+ C-Exp.} instead enables patch-correction experience to guide the revision of failed patches. The full \tool{} enables contrastive analysis and both experience tracks.

Each configuration is run three times. Before each run, the accumulated skill base is reset to the same initial state, and the 200 instances are processed in the same fixed order. For each configuration, the per-instance mechanism analyses below use the median run by resolved instances. Unless otherwise stated, each enabled experience track retrieves at most two entries.

\begin{table}[t]
  \centering
  \footnotesize
  \setlength{\abovecaptionskip}{0.1cm}
\caption{Repair success conditioned on contrastive-pair quality.}
  \label{tab:rq2-conditional}
  \setlength{\tabcolsep}{3pt}
  \renewcommand{\arraystretch}{1.08}
  \begin{tabularx}{\linewidth}{@{}Y C{0.27\linewidth} C{0.27\linewidth}@{}}
    \toprule
   \textbf{Configuration}      & \makecell[c]{$P(\mathrm{resolved}\mid$\\$\mathrm{Ideal})$}      & \makecell[c]{$P(\mathrm{resolved}\mid$\\$\mathrm{No Ideal})$} \\    \midrule
    Contrastive analysis
      & 126/156 (80.77\%)
      & 24/44 (54.55\%) \\
    + M-Exp.
      & 140/168 (83.33\%)
      & 19/32 (59.38\%) \\
    + C-Exp.
      & 144/157 (91.72\%)
      & 30/43 (69.77\%) \\
    \tool{} (Full)
      & \textbf{160/168 (95.24\%)}
      & \textbf{24/32 (75.00\%)} \\
    \bottomrule
  \end{tabularx}
  \vspace{-0.25cm}
\end{table}
\paragraph{Component effects}
Across the three runs, the ordering of the five configurations is unchanged; all sample standard deviations are at most 0.9 percentage points, and every
within-configuration range is at most 1.5 points. On average, contrastive analysis raises the resolution rate from 48.0\% to 74.8\% (+26.8 points), while the full system reaches 91.8\%, a further 17.0-point improvement.

Across the three runs, M-Exp. adds 2.5--5.5 points and C-Exp. adds 12.0--13.0 points across the two settings of the other track; the corresponding interaction term remains between 0 and 0.5 points. This consistency indicates
that the two experience types provide largely distinct and complementary information.

Table~\ref{tab:rq2-conditional} groups instances by contrastive-pair quality. \textit{Ideal} means that the original PoC, a same-crash variant, and an oracle-passing variant reach a common probe site; \textit{No ideal} denotes cases where this condition is not met, such as when all variants still crash.

\paragraph{Mutation experience improves the evidence available to later stages}
Table~\ref{tab:rq2-conditional} separates instances with ideal contrastive pairs from fallback instances. Across all configurations, ideal-pair instances are repaired more often than fallback instances, confirming that the quality of
the contrastive evidence strongly affects downstream repair.

M-Exp. increases the number of ideal pairs from 156 to 168 and reduces fallback cases from 44 to 32. Its effect is therefore mainly indirect: it helps the Mutator preserve PoC structure while crossing the failure boundary, giving the Analyzer clearer evidence for RepairSpec inference. The gain is concentrated
in structured binary inputs: ideal pairs increase from 35 to 40 for Image \& Graphics and from 44 to 50 for Media Containers, while the Other category remains unchanged at 36. These formats require mutations that preserve framing and parser reachability.

C-Exp. has a different effect. It barely changes ideal-pair construction ($156\rightarrow157$), but improves success within both ideal and fallback groups. This matches its intended role: M-Exp. improves the evidence available to the Analyzer, whereas C-Exp. improves the Patcher's ability to recover from failed candidate patches.

\paragraph{Correction experience supports feedback-driven recovery}
The full system repairs 112 instances in Loop~1, 69 additional instances in Loop~2, and 3 in Loop~3, reaching 184/200 (92.0\%). The latter 72 instances produce 72 correction trajectories, one per instance recovered after at least one failed patch. Extending the loop to five rounds yields no further repairs, indicating that the remaining failures require different evidence, invariants, or repair scopes rather than additional iterations.

Varying the per-track retrieval budget over $k\in\{1,2,3,4\}$ yields 168, 184, 180, and 178 repairs, respectively, supporting $k{=}2$. Because correction retrieval ranks failed-diff and validation-feedback similarity, larger budgets necessarily add lower-ranked trajectories. The decline for
$k{>}2$ suggests that examples involving different repair locations or invariants can distract the Patcher, particularly when recovery depends on a small but decisive edit such as moving a guard upstream or extending it to sibling paths.

\vspace{3mm}

\begin{custommdframed}

\textbf{Answer to RQ2:} Contrastive analysis provides the main gain by turning a raw failure report into execution-grounded boundary evidence, which substantially improves downstream patch generation. The two experience tracks then improve different stages of the loop: M-Exp. helps construct more ideal contrastive pairs, whereas C-Exp. helps recover from failed candidate patches. Thus, the ablation shows that \tool{}'s effectiveness comes from a staged repair process: first obtaining better repair evidence, then using accumulated experience to refine how that evidence is constructed and applied.

\end{custommdframed}

\begin{table*}[t]
  \centering
  \tabledoublesetup
  \setlength{\abovecaptionskip}{0.1cm}
  \caption{Performance comparison on PatchEval across Go, Python, and JavaScript. Values in parentheses denote resolution rates.}
  \label{tab:rq3-patcheval}
  \setlength{\tabcolsep}{5pt}
  \renewcommand{\arraystretch}{1.12}
  \begin{tabularx}{\textwidth}{@{}L{0.22\textwidth} *{5}{Z}@{}}
    \toprule
    \textbf{Method}
      & \textbf{Go}
      & \textbf{Python}
      & \textbf{JavaScript}
      & \textbf{Total}
      & \textbf{Avg.\ Cost} \\
    \midrule
    SWE-agent~\cite{yang2024sweAgent}
      & 28 (33.7)
      & 17 (25.8)
      & 33 (43.4)
      & 78 (34.7)
      & \$2.39 \\
    OpenHands~\cite{wang2024openhands}
      & 32 (38.6)
      & 18 (27.3)
      & 35 (46.1)
      & 85 (37.8)
      & \$4.64 \\
    Claude Code~\cite{claudecode2025}
      & 33 (39.8)
      & 16 (24.2)
      & 32 (42.1)
      & 81 (36.0)
      & \$6.40 \\
    \midrule
    \textbf{\tool{}}
      & \textbf{67 (80.7)}
      & \textbf{51 (77.3)}
      & \textbf{48 (63.2)}
      & \textbf{166 (73.8)}
      & \textbf{\$0.56} \\
    \bottomrule
  \end{tabularx}
  \vspace{-0.3cm}
\end{table*}

\subsection{RQ3: Generalizability}\label{sec:rq3}

On PatchEval's 225 Go, Python, and JavaScript instances, \tool{} uses language-specific test feedback and retrieves experience by CWE and language, while retaining the same controller and contrastive workflow. Table~\ref{tab:rq3-patcheval} compares it with GPT-5-based baselines.

\tool{} resolves 166 out of 225 instances (\textbf{73.8\%}) at an average cost of \$0.56 per task, nearly doubling the resolution rate of the strongest baseline (OpenHands, 37.8\%) at a fraction of the per-task cost. This result is consistent with the framework-level advantage observed on SEC-Bench, where same-backbone comparisons isolate the effect of the repair workflow. Per-language rates are \textbf{80.7\%} on Go (67/83), \textbf{77.3\%} on Python (51/66), and \textbf{63.2\%} on JavaScript (48/76). Although lower than \tool{}'s 92.0\% rate on SEC-Bench, the 73.8\% overall PatchEval rate demonstrates that the contrastive runtime analysis and skill accumulation mechanisms transfer effectively from C/C++ memory safety to diverse languages and vulnerability classes.

JavaScript exhibits a noticeably lower resolution rate, primarily due to its complex build toolchains (Babel, Webpack, custom preload scripts) that hinder both variant generation and build-graph-preserving edits. We inspected all 59 PatchEval failures across the three languages and found that 26 generated patches are logically consistent with the ground-truth developer fixes but fail benchmark validation for non-repair reasons. These include \textbf{test-suite mismatches} (10 cases), where legacy regression tests enforce pre-fix behavior, such as expecting malformed input to be silently accepted; and \textbf{benchmark environment issues} (16 cases), where even the ground-truth developer patch fails due to proxy or Docker image mismatches. Diff comparisons and logs for these 26 cases are included in the replication package.

\vspace{3mm}
\begin{custommdframed}
\textbf{Answer to RQ3:} \tool{} generalizes beyond C/C++ on PatchEval, resolving 166/225 instances (73.8\%) across Go, Python, and JavaScript. It nearly doubles the strongest baseline, OpenHands (37.8\%), at a lower average cost of \$0.56 per task, with per-language rates of 80.7\% on Go, 77.3\% on Python, and 63.2\% on JavaScript.
\end{custommdframed}

\section{Discussion}
\label{sec:discussion}

\subsection{Failure Modes}
\label{sec:failure-modes}

We manually inspect the SEC-Bench outcomes to understand where \tool{} still fails. As reported in Section~\ref{sec:rq1}, 77 benchmark-validated patches are not semantically correct, including 44 incomplete/PoC-specific and 33 behavior-regressive patches. \tool{} also fails to produce any benchmark-validated patch for 16 instances, including 9 root-cause tracing failures and 7 incomplete candidate repairs. This section gives representative cases.

\paragraph{Root-cause tracing gaps}
This failure mode occurs when the crash site is only where an invalid state is later used, while the real fix must be placed where that state is created or should have been rolled back. In gpac.cve-2023-41000, \tool{} follows a late command-buffer lifetime symptom, but the developer patch repairs an earlier rollback rule in \texttt{BM\_ParseCommand}.

\paragraph{Incomplete or PoC-specific repairs}
In gpac.cve-2021-32439, \tool{} protects the observed stbl\_AppendSize path. Manual inspection labels the patch incomplete because the same packed-sample invariant is shared by several sibling sample-table operations. Listing~\ref{lst:gpac-poc-specific} shows the pattern. The patch blocks the path exercised by the PoC, but other functions that update the same sample-table state remain unprotected.

\begin{lstlisting}[escapeinside={|}{|}, caption={gpac.cve-2021-32439.}, label={lst:gpac-poc-specific}, language=C]
/* Covers only the PoC-exercised path. */
int stbl_AppendSize(...) {
    if (!valid_packed_sample_state(stbl))
        return ERROR;
    append_size(stbl, ...);
}
/* Sibling paths require the same guard. */
int stbl_AppendChunk(...) {
     |\textcolor{red}{    append\_chunk(stbl, ...);   // need same guard}|
}
int stbl_UpdateSample(...) {
    |\textcolor{red}{    update\_sample(stbl, ...);  // need same guard}|
}
\end{lstlisting}

This case shows that after a candidate condition is found, the repair system should search for other functions that update the same fields or enforce the same file-format rule.

\paragraph{Behavior regressions}
In openjpeg.cve-2016-7445, a PNM header carries three fields---width, height, and maxval---that a legal file may place either on a single line or split across multiple lines. To block the malicious PoC, \tool{} adds a check that rejects any input whose three fields are not all on the current line. This stops the attack, but it also wrongly rejects legal multi-line headers, treating valid input as if it were an attack. Listing~\ref{lst:openjpeg-regression} shows the over-strict guard and the legal path it wrongly blocks (both in red).


\begin{lstlisting}[
caption={openjpeg.cve-2016-7445.},
label={lst:openjpeg-regression},
language=C,
escapeinside={(*@}{@*)},
basicstyle=\ttfamily\footnotesize,
columns=flexible,
keepspaces=true,
breaklines=true,
breakatwhitespace=false
]
/* Over-strict guard: requires all three fields on one line. */
(*@\textcolor{red}{if (!current\_line\_contains\_all\_header\_fields())}@*)
(*@\textcolor{red}{\hspace*{2.5em}return ERROR; /* rejects valid headers */}@*)
read_next_token(&width);
read_next_token(&height);
read_next_token(&maxval);
\end{lstlisting}

Such cases show that vulnerability repair needs positive boundary tests in addition to malicious inputs, so a patch is accepted only when it removes the vulnerability without rejecting valid behavior.

\subsection{Implications}

These outcome groups require different improvements. Root-cause tracing failures suggest probing where invalid state is created, including parser failure and rollback boundaries, rather than only the final crash site. Oracle-passing incomplete repairs require searching sibling functions that manipulate the same fields or implement the same format rule. Behavior regressions require positive boundary tests in addition to malicious inputs, so that a patch is accepted only when it removes the vulnerability without rejecting valid behavior. The PatchEval mismatches further show that execution-based benchmarks can be affected by stale regression expectations and environment drift, so we report these cases separately rather than counting them as successful repairs.

\section{Threats to Validity}
\label{sec:threats}

\subsection{Internal Validity}

Experience reuse may introduce order dependence or leakage across instances. We mitigate this risk by resetting the accumulated skill base before each repetition, using the same instance order across configurations, excluding same-CVE entries, and retrieving only experiences whose vulnerability disclosure predates the current instance. The reported three-run results therefore measure stochastic variation under a fixed experience curriculum rather than all possible instance orders. Pretraining contamination in proprietary LLMs cannot be fully ruled out, but \tool{}'s 29.5--42.0 point gains over same-backbone GPT-5-mini baselines and its DeepSeek-V3.2 result make backbone memorization unlikely to explain the relative improvements.

\subsection{Construct Validity}

\tool{} derives each RepairSpec from a finite set of generated executions, so the inferred relation may not cover the full input space. We therefore treat a RepairSpec as an empirically grounded repair hypothesis rather than a formal invariant, and every generated patch must still pass build, vulnerability-oracle, and regression-oracle checks. These benchmark oracles are not complete semantic specifications. To avoid equating oracle passing with correctness, we separately perform a manual semantic review of all benchmark-validated SEC-Bench patches and report semantic correctness apart from benchmark validation.

\subsection{External Validity}

The evaluation covers 425 instances across C/C++, Go, Python, and JavaScript, but does not represent all vulnerability classes, software architectures, or deployment environments. In particular, vulnerabilities involving concurrency, cryptography, distributed state, or proprietary code may require evidence and repair mechanisms beyond those evaluated here.
Porting \tool{} to a new language mainly requires implementing language-specific wrappers for execution; the gated workflow remains unchanged. The reported cross-language results therefore support portability across the evaluated ecosystems rather than language-independent generality.

\section{Related Work}
\label{sec:related}

\subsection{Repair Constraints and Vulnerability Repair}

Traditional APR repairs functional failures specified by tests. Template-, search-, and learning-based methods~\cite{xuan2016nopol,huang2019using,
liu2019tbar,fu2022vulrepair,jiang2021cure} either instantiate edit patterns or learn fixes from historical patches, often with the repair region already
localized. Constraint-based repair, represented by SemFix~\cite{SemFix}, derives symbolic repair constraints from test executions. \tool{} also uses an
intermediate condition to guide patch generation, so repair-constraint construction itself is not our novelty. The difference is that \tool{} derives an empirical RepairSpec from aligned failure-reproducing and oracle-passing executions, records concrete boundary evidence and patch scope, and passes this specification to a repository-level LLM patcher.

Vulnerability-oriented analysis has studied neighboring exploit inputs and execution differences. VulnLoc~\cite{VulnLoc} localizes suspicious code from
exploit-neighborhood fuzzing, CrashRepair~\cite{CrashRepair} infers crash-free
specifications with concolic execution, and delta debugging and cause-effect
isolation~\cite{zeller2002simplifying,10.5555/3489212.3489226} reduce failure inputs or identify outcome-relevant state changes. LLM-based AVR systems further
use CWE knowledge, retrieved fixes, iterative feedback, sanitizer diagnostics, or validation constraints~\cite{DBLP,liu2024crepair,
dong2025infcodecintentguidedsemanticretrieval,cao2025enhancing, xia2024automated,yu2025patchagent,kim2025SAN2PATCH,zhang2026vulnresolver}. We therefore do not claim exploit-neighborhood analysis, differential execution,
or repair constraints individually. \tool{} combines them in an AVR workflow that aligns failure-reproducing and oracle-passing executions at shared probe sites, encodes the observed separating relation as a RepairSpec, and validates repository-level patches across C/C++, Go, Python, and JavaScript.

\subsection{Experience Accumulation in LLM Agents}

Voyager~\cite{wang2023voyager} stores executable skills, Reflexion
~\cite{shinn2023reflexion} stores textual feedback, and experiential
co-learning~\cite{qian2024experiential} shares task-level lessons among software-engineering agents. Repository-level repair systems reuse similar issues \cite{zhao2025recodeimprovingllmbasedcode} or patches~\cite{zhang2024autocoderover}; ExpeRepair~\cite{ExpeRepair} stores repair demonstrations and semantic insights from prior issue-driven repairs. These memories improve reuse but do not represent the PoC-construction step required by exploit-driven AVR.

AVR requires two distinct reusable decisions. A useful mutation must preserve the input format and vulnerable path while crossing the execution boundary; patch refinement must identify how an edit using the wrong variable, location,
boundary, or repair scope was corrected. Prior systems may accumulate issue-level repair experience, but they do not explicitly represent exploit-driven PoC construction or activate failed-patch correction trajectories after a rejected AVR candidate. Agenticrepair~\cite{AgentMem} accumulates interaction traces online, but does not separately represent PoC-boundary construction and failed-patch correction.

\tool{} stores these decisions as records with different contents and activation points. A mutation record contains the PoC format, applied transformation, and observed same-crash/oracle-passing outcomes and is retrieved before variant generation. A correction trajectory contains each rejected diff, its compiler or oracle feedback, and the final benchmark-validated diff and is retrieved only after the current Patcher produces a failed candidate. This separation lets experience directly guide both contrastive-evidence
construction and patch refinement.

\section{Conclusion}
We presented \tool{}, an LLM-based automated vulnerability repair framework that addresses two key limitations of existing approaches: the lack of repair-useful boundary evidence from a single failing execution and the limited reuse of cross-instance AVR experience. By generating boundary-straddling PoC variants and executing them under aligned state probes, \tool{} identifies candidate state relations that distinguish failure-reproducing and oracle-passing executions. A dual-track skill base separately accumulates mutation and patch-correction experience for future reuse. On SEC-Bench and PatchEval, \tool{} achieves 92.0\% and 73.8\% resolution rates, outperforming all evaluated baselines while maintaining competitive cost. Ablation studies further confirm complementary contributions from contrastive analysis and skill accumulation.
Overall, our framework provides a practical and effective paradigm for advancing LLM-based automated vulnerability repair.


\bibliographystyle{ACM-Reference-Format}

\bibliography{ref}


\end{document}